\documentstyle[epsfig]{aipproc}

\def\f2cc{$F_2^{c\bar{c}}$}
\begin{document}
\title{ 
Heavy Quark Production at HERA}
\author{
Andreas~B.~Meyer\footnote{Contribution to ``Workshop on Heavy Quarks at Fixed
Target'', Centro Brasileiro de Pesquisas Fisicas, 
Rio de Janeiro, Brazil, Oct.\,9 2000 }    \\
(on behalf of the H1 and ZEUS Collaborations)}
\address{University of Hamburg\\ II. Institut f\"ur Experimentalphysik\\
Luruper Chaussee 149\\D 22761 Hamburg\\
E-mail:Andreas.Meyer@desy.de}
\maketitle

\baselineskip=11.6pt
\begin{abstract}
Recent results on heavy flavor production at HERA are presented. New
measurements of charm production are available in the region of
$ Q^2 <$ 1000 GeV$^2$ for $D^*$ and in the photoproduction
limit, $Q^2\rightarrow 0$, for $D_s$. The 
results are compared to theoretical calculations and the charm 
proton structure function \f2cc$(x,Q^2)$ is extracted.
A new independent measurement of the $b\bar{b}$ cross section using
a $b$-lifetime tag is reported.
\end{abstract}
\baselineskip=14pt
\section{Introduction}
At the storage ring HERA positrons of
27.5 GeV collide with protons of 920 GeV 
resulting in a center of mass energy $\sqrt{s}=320$ 
GeV. 
The event kinematics of $ep$ scattering is described by 
$s=k+P$, $Q^2=-(k-k')^2$, $y=q \cdot P / k \cdot P$, $x=Q^2/sy$ and $W^2_{\gamma
p}=(q+P)^2=sy-Q^2$ where $k$, $k'$, $P$ and $q$ are the 4-vectors of 
the incoming/outgoing electron, incoming proton and exchanged photon, respectively.
\begin{figure}[t] 
\unitlength1.0cm
\begin{picture}(12.,6.0)
\put(0.0,5.0){a)}
\put(8.5,4.5){b)}
\put(1.0,0.5){\epsfig{file=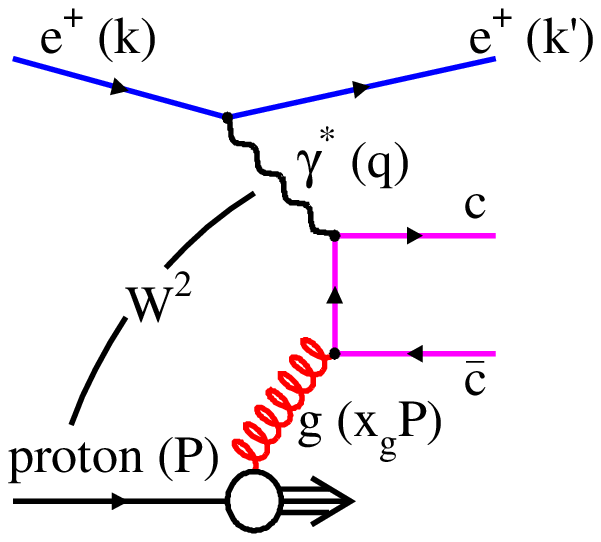,width=5.cm}}
\put(7.,-0.2){\epsfig{file=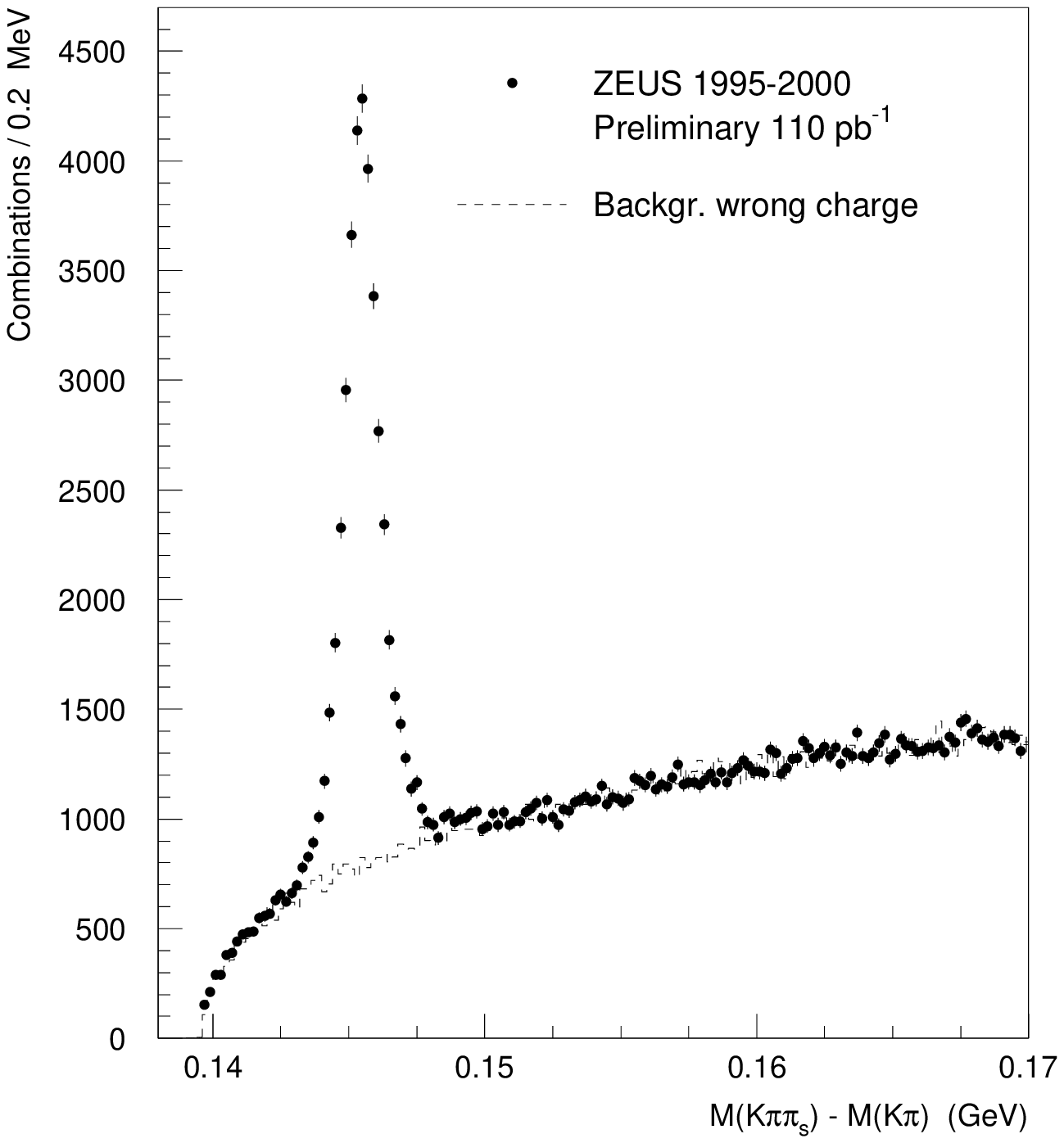,width=7.9cm,height=6.2cm}}
\end{picture}
\caption{\it 
a) Charm Production via Boson-Gluon-Fusion; 
b) Mass difference distribution $\Delta M =M_{K\pi\pi_s}-M_{K\pi}$ as obtained
from 110 pb$^{-1}$ by ZEUS.
 \label{fig:1} }
\end{figure}

\subsection{Heavy Flavor Production Mechanism}
Heavy quark production predominantly occurs
via the process of boson gluon fusion (BGF) sketched in 
fig.\ref{fig:1}a. It is thus directly
sensitive to the gluon distribution in the proton. At HERA the 
measurement of heavy flavor processes is a powerful means to study 
perturbative QCD down to values of $x \ge 10^{-4}$. 
In particular it can be tested if the parton distributions 
are universal, i.e.\,independent of the nature of the 
hard scattering process. The presence of two large scales, the squared 
4-momentum transfer $Q^2$ and the heavy quark masses $m_{c,b}^2$
makes heavy flavor production a useful test bed for the study of 
perturbative QCD resummation techniques.

Several schemes to calculate heavy quark cross sections in next-to-leading order 
(NLO) are available. The program HVQDIS\cite{hvqdis} implements a fixed 
flavor scheme at fixed order $\alpha_s^2$, assuming that the proton is made up of 
three active quark flavors (uds). The parton distributions are obtained 
by the DGLAP equation\cite{dglap}. The heavy quark pairs are produced at 
the perturbative level via BGF. The results of HVQDIS are expected
to be most accurate for $Q^2 \sim m_c^2$. At large $Q^2$ terms of orders 
higher than $\alpha^2_s$ contain log$(Q^2/m_c^2)$ factors that can become large.
The CCFM equation\cite{ccfm} approximates the parton 
cascade as emitting gluons in an angular ordered way. The cross 
section is calculated according to the $k_t$-factorization theorem by
convoluting an unintegrated gluon density with an off-shell BGF matrix element. 
A new hadron level Monte Carlo generator CASCADE\cite{cascade} is available which 
implements the CCFM equation.

\begin{figure}[b] 
\unitlength1.0cm
\begin{picture}(12.,5.0)
\put( 6.0,4.3){a)}
\put(13.0,4.3){b)}
\put(0.5,-0.5){\epsfig{file=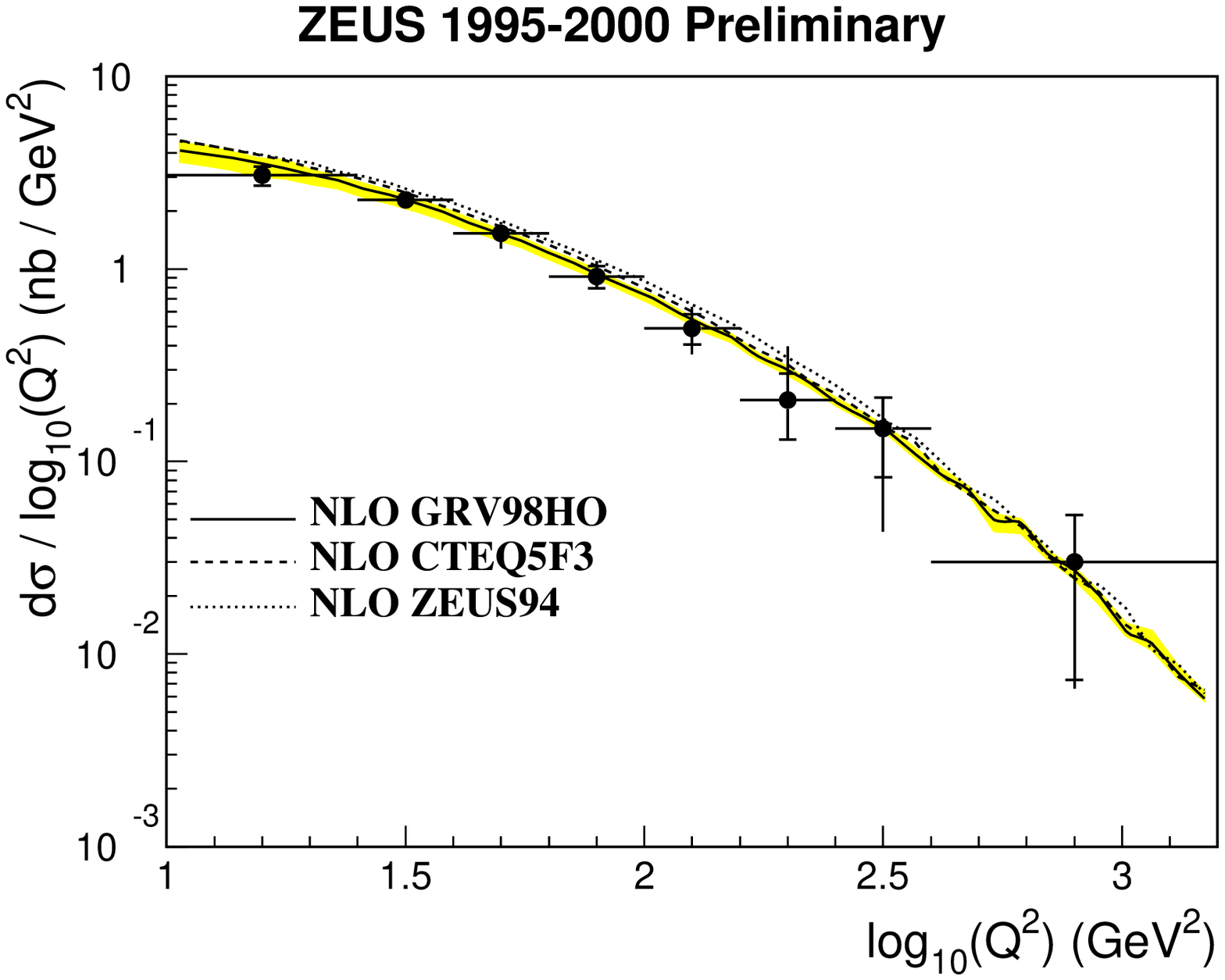,width=5.5cm}}
\put(7.5,-0.5){\epsfig{file=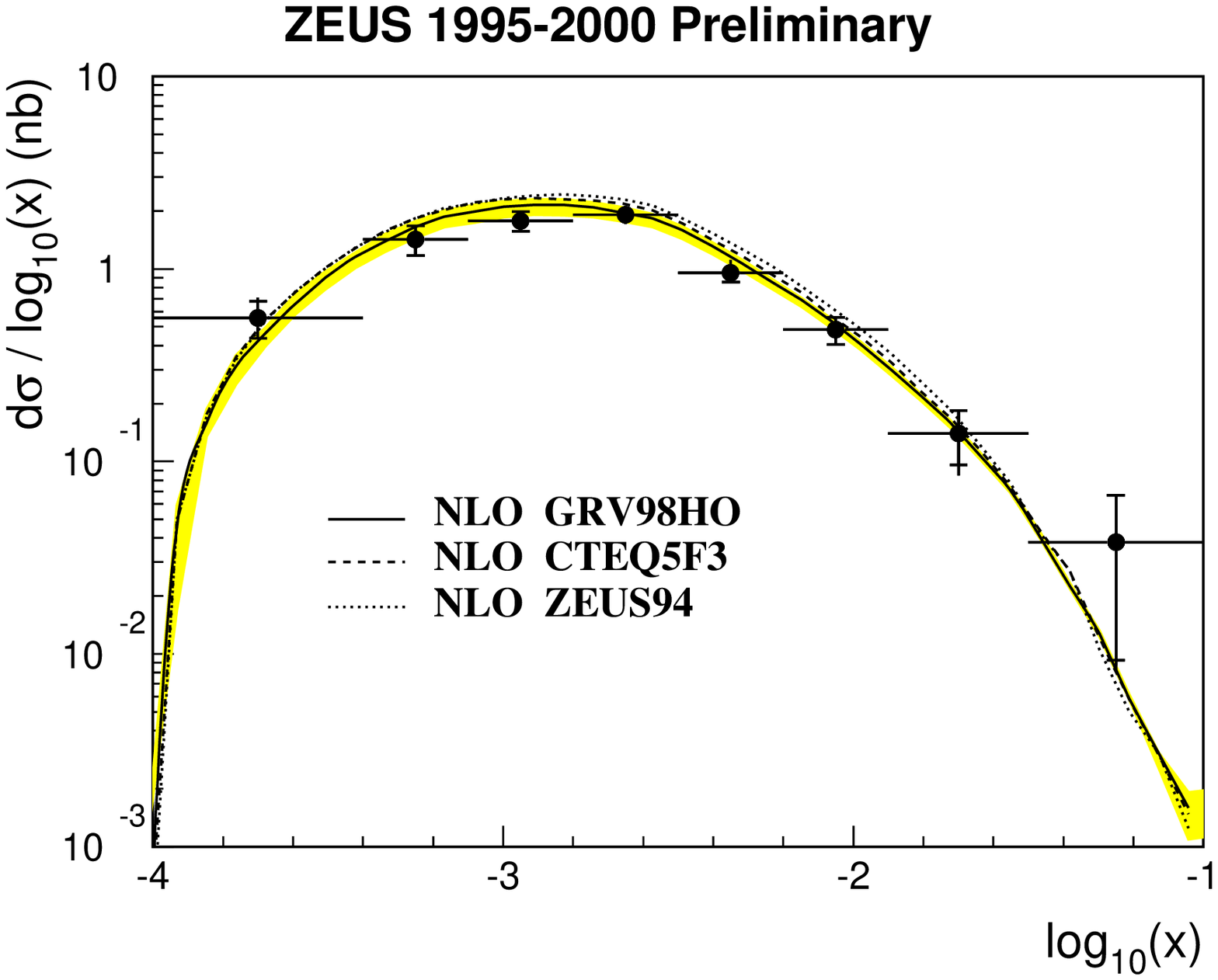,width=5.5cm}}
\end{picture}
\caption{\it Differential $D^*$ production cross sections as a function of $Q^2$
(a) and $x$ (b) compared to the NLO QCD calculation of HVQDIS with different
parton density distributions.  
  \label{fig:dstarhighq} }
\end{figure}

\subsection{Experimental Procedure}
Charm events are tagged by
reconstructing $D^*$-mesons (H1\cite{osakaa}, ZEUS\cite{osakab,zeusf2cc}), 
$D_s$ mesons (ZEUS\cite{dsubs}) or semi-leptonic decays into electrons 
(ZEUS\cite{osakac}).
Visible cross sections of $D$-meson production are measured in 
the central region of detector acceptance, at $p_{\perp}^{D} > 1.5 $ GeV/c 
and $|\eta| < 1.5$, where
$p_{\perp}$ is the transverse momentum of the $D$-meson and $\eta$
denotes the pseudorapidity. 
Measurements of the $b\bar{b}$-cross section have been 
performed through semi-leptonic decays in muons (H1\cite{bh1}, 
ZEUS\cite{bzeustamp}) or 
electrons (ZEUS\cite{bzeus}) and, more recently, by exploiting 
the long lifetime of $B$-hadrons (H1\cite{osakae}).

Since the beginning of HERA operation in 1992 the luminosity 
has continuously increased, yielding a total integrated luminosity 
collected by each of the two collider experiments H1 and ZEUS of more
than 100 pb$^{-1}$.
Fig.\ref{fig:1}b shows the $D^*-D^0$ mass difference distribution measured by 
the ZEUS experiment. There are roughly 27,000 events in the
signal illustrating the excellent capabilities of HERA for the study of the
charm production mechanism and also for charm meson spectroscopy studies\cite{osakad}.
In winter 2000/2001 the HERA accelerator is being upgraded. In the future it
is expected to deliver a yearly integrated luminosity of 150 pb$^{-1}$.

\section{Open Charm Production}

\begin{figure}[t] 
\unitlength1.0cm
\begin{picture}(12.,5.5)
\put( 5.7,4.4){a)}
\put(12.1,4.4){b)}
\put(1.,0.){\epsfig{file=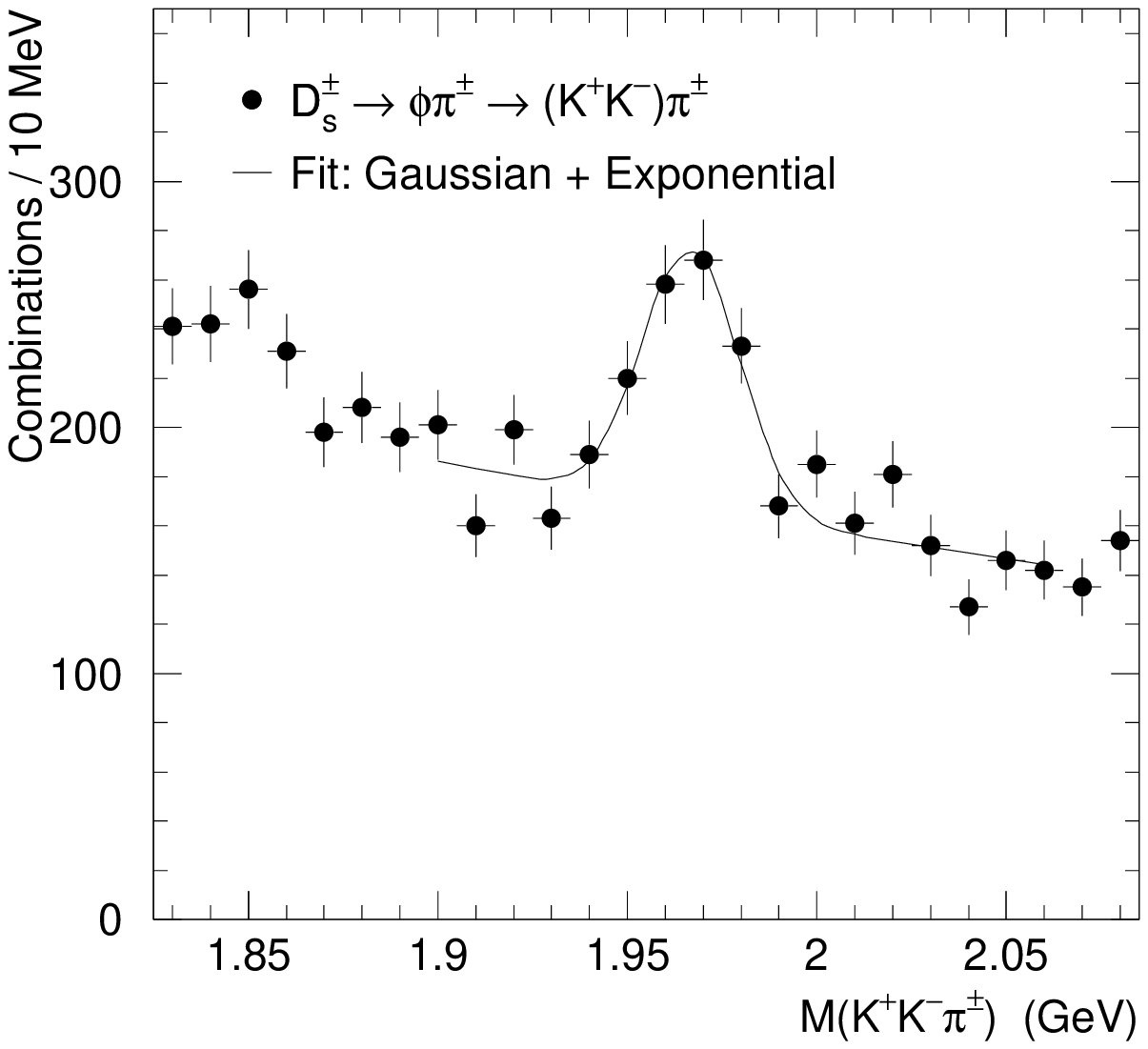,width=5.5cm}}
\put(7.5,0.){\epsfig{file=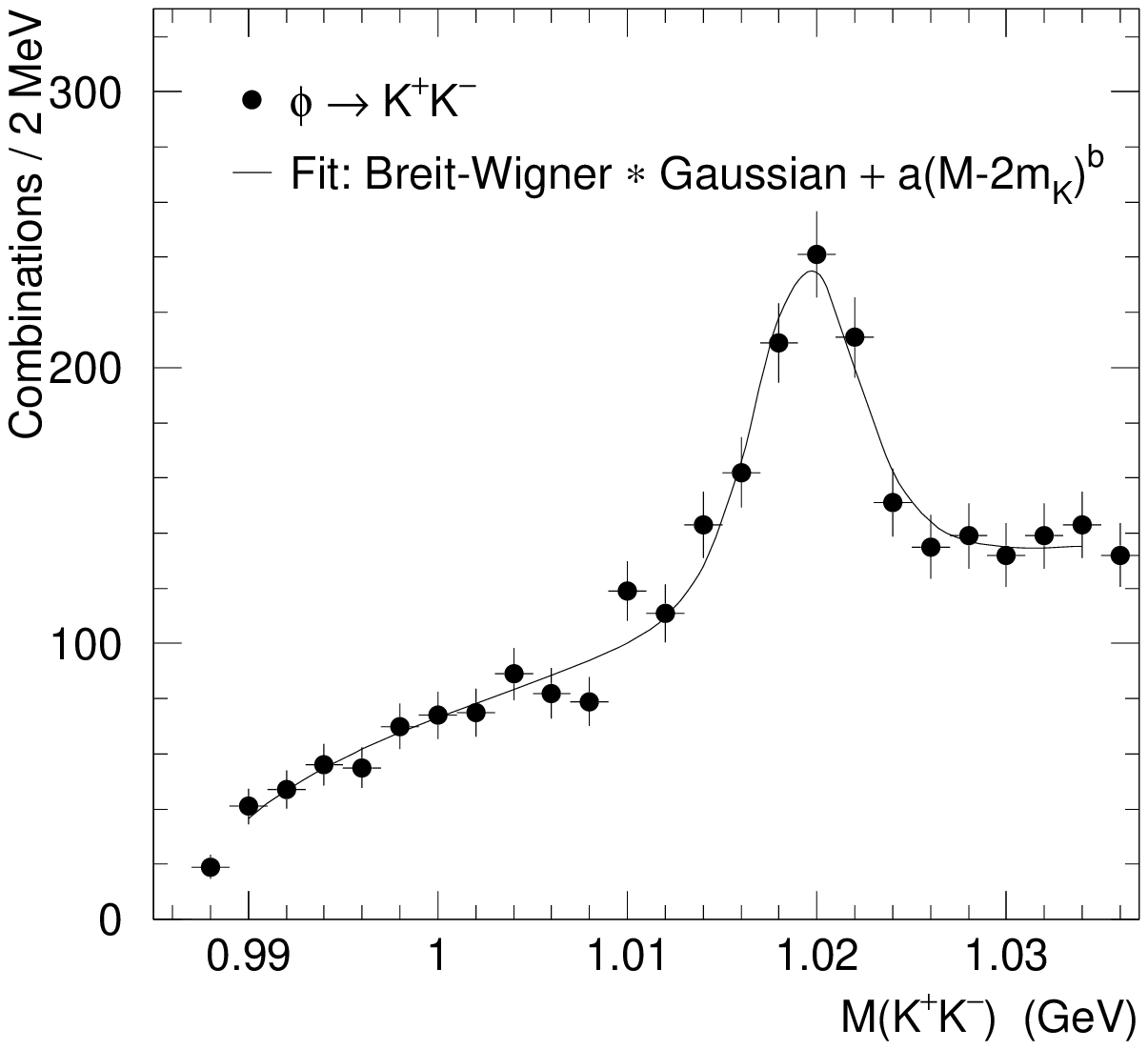,width=5.5cm}}
\end{picture}
\caption{\it a) 
Distribution of $M(K^+K^-\pi^\pm)$ for events inside 
the $\phi$ mass range $1.0115 < M(K^+K^-) < 1.0275$ GeV as measured by ZEUS;
b) $M(K^+K^-)$ distribution for events inside the $D_s$ mass range.
\label{fig:dsubssignal} }
\end{figure}

\subsection{$D^*$ production at high $Q^2$}
Fig.\ref{fig:dstarhighq} shows the results of 82.6 pb$^{-1}$ of $D^*$
data collected by the ZEUS experiment
\cite{osakab} in the region $10 < Q^2 < 1000$ GeV$^2$.
The shaded band represents the theoretical calculation 
as implemented in the HVQDIS program.  
Here the charm quark mass is varied between 1.3 and 1.6 GeV. 
The input gluon distribution is derived from results of different
QCD fits to the inclusive proton structure 
function\cite{zeusqcd,grv98,cteq5}.
The Peterson fragmentation function\cite{peterson}
is used to describe the probability that 
a $c$-quark hadronizes into a $D^*$ meson.
The theoretical predictions agree well with the data
indicating that, within the statistics presently available, 
the BGF calculation in the fixed flavor scheme of order $\alpha_s^2$
fully accounts for the charm cross section up to
very high values of $x$ and $Q^2$.

\subsection{$D^*$ and $D_s$ production in photoproduction}

\begin{figure}[t] 
\unitlength1.0cm
\begin{picture}(12.,5.5)
\put(5.7,4.2){a)}
\put(12.7,4.2){b)}
\put(-2.,-5.5){\epsfig{file=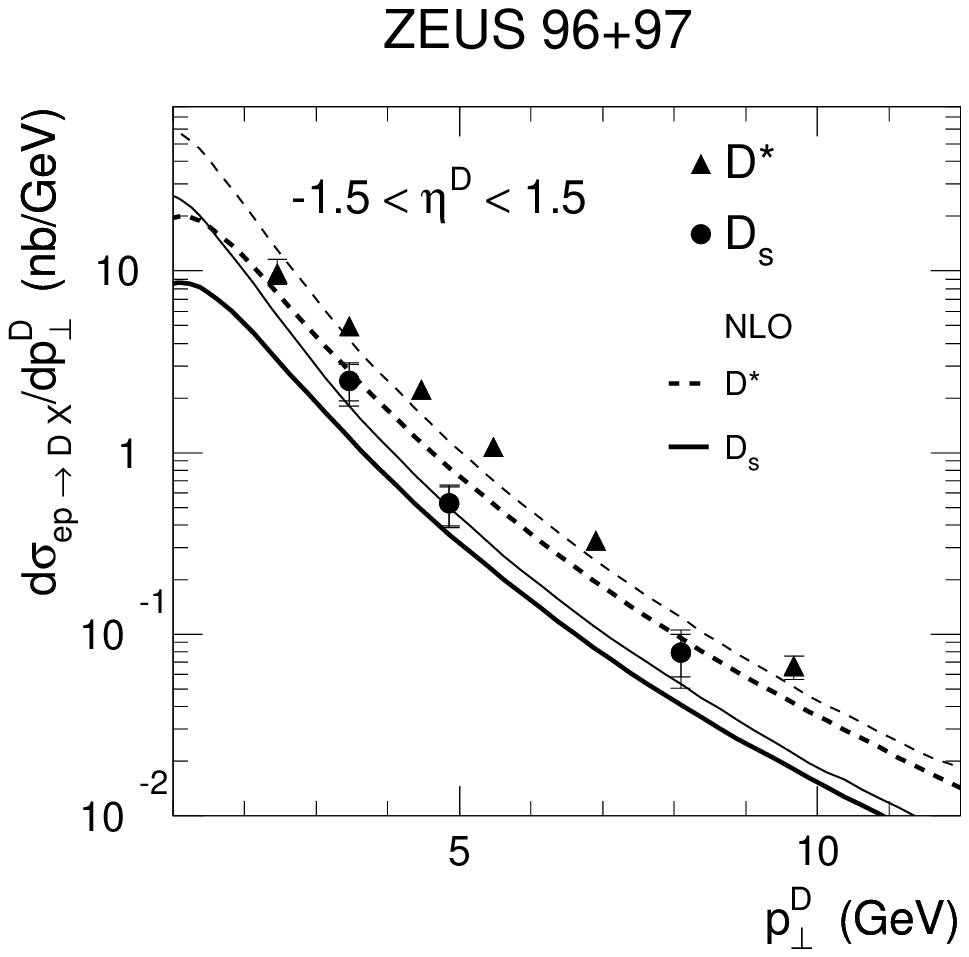,width=7.0cm}}
\put( 5.,-5.5){\epsfig{file=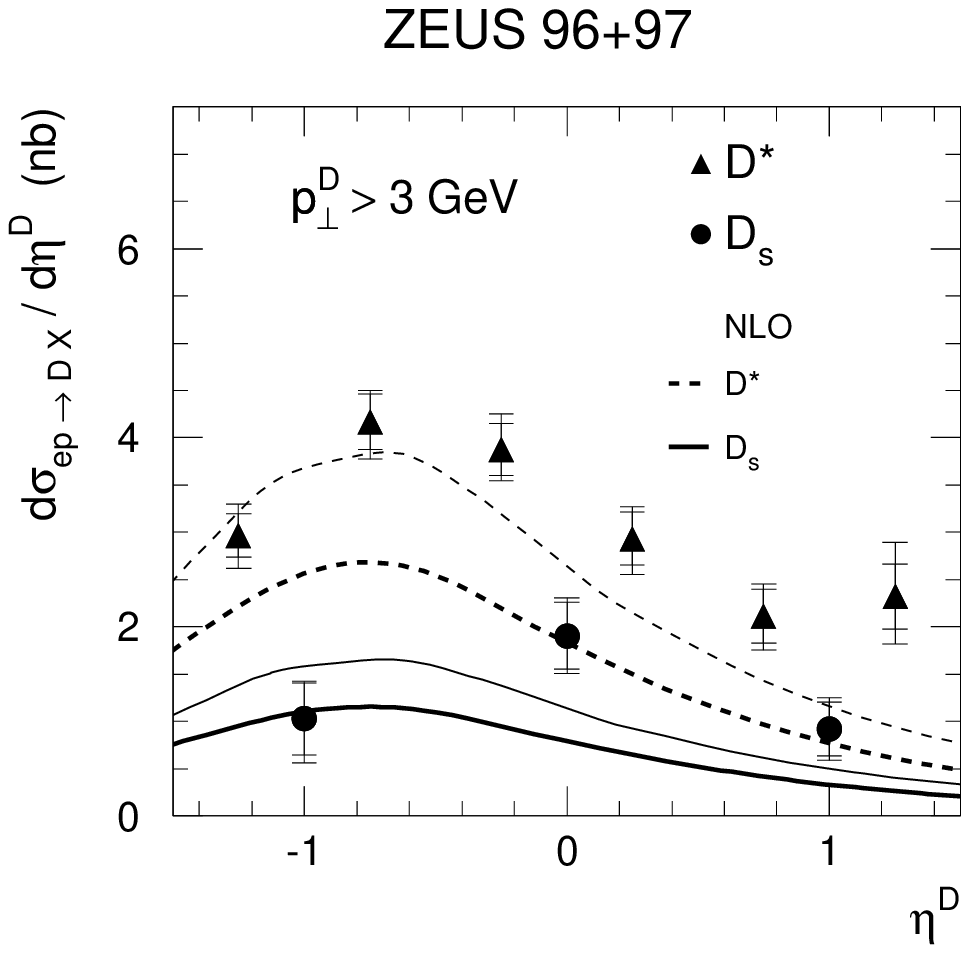,width=7.0cm}}
\end{picture}
\caption{\it Differential cross section 
(a) $d\sigma/dp_{\perp}^D$ and (b) $d\sigma/d\eta^D$
for $D^*$ (triangles) and $D_s$ (bullets) in comparison with 
NLO calculations (see text).
\label{fig:dsubs} }
\end{figure}

\begin{figure}[t] 
\unitlength1.0cm
\begin{picture}(12.,10.)
\put(1.,0.){\epsfig{file=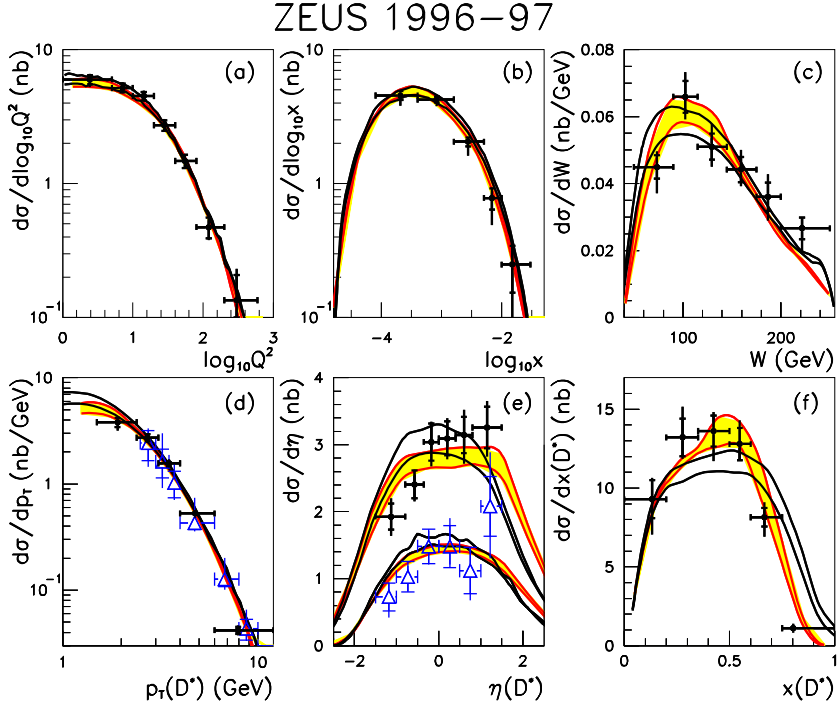,height=10cm}}
\end{picture}
\caption{\it Differential cross sections for $D^*$ production from the $K\pi\pi$
final state (bullets). Fig.~d and e also show the cross section for the 
$K4\pi$ decay channel.
 \label{fig:zeusdstarsmallq} }
\end{figure}

\begin{figure}[t] 
\unitlength1.0cm
\begin{picture}(12.,4.5)
\put(3.8,3.9){a)}
\put(9.0,3.9){b)}
\put(14.2,3.9){c)}
\put(-0.2,0.){\epsfig{file=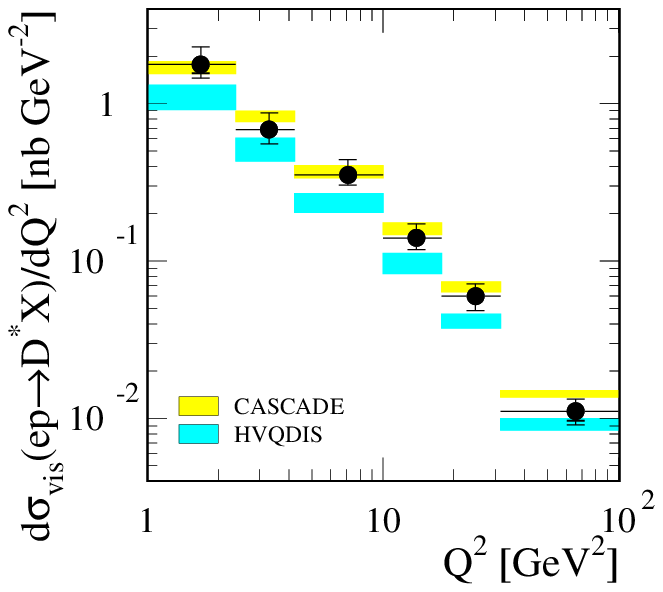,height=4.4cm}}
\put(5.,0.){\epsfig{file=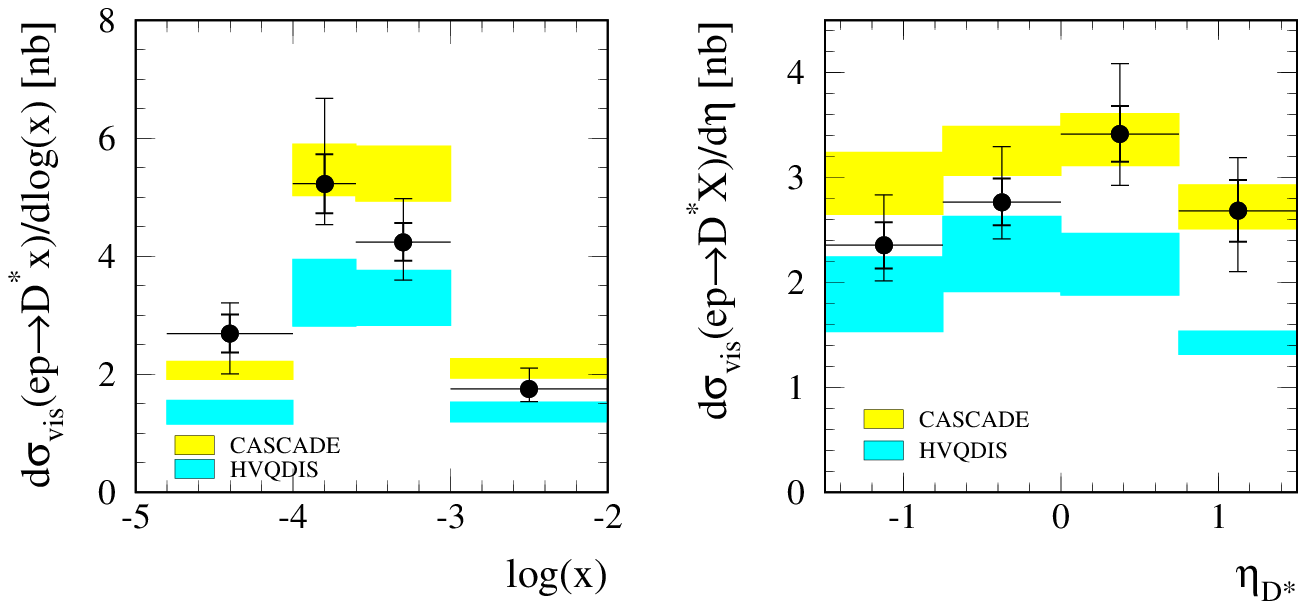,height=4.6cm}}
\end{picture}
\caption{\it Single differential inclusive cross section $\sigma(ep\rightarrow eD^{*}X)$
as a function of a) $Q^2$, b) $x$, c) $\eta$. The expectations from HVQDIS (CASCADE) 
are indicated by the dark (light) shaded bands. 
 \label{fig:h1dstarsmallq} }
\end{figure}

Visible cross sections of $D^*$ mesons in photoproduction 
have been presented before\cite{dstargp}.
More recently a measurement of $D_s$-mesons was performed
using the decay $D_s \rightarrow \phi \pi$ and $\phi \rightarrow KK$.
Fig.\ref{fig:dsubssignal} shows the signals of the reconstructed $D_s$ and 
the corresponding $\phi$\cite{dsubs}. 
The visible cross section is extracted in the region 
$ 12 > p_{\perp}^{D^*} > 3$ GeV and
$ | \eta^{D^*} | < 1.5 $ yielding 
$\sigma^{vis}_{D_sX} = (3.79 \pm 0.59 (stat) ^{+0.26}_{-0.46} 
(syst) \pm 0.94 (BR))$ nb  where the last error accounts 
for the uncertainty of the branching ratio. 
Fig.\ref{fig:dsubs} compares the measured visible 
production cross sections to NLO predictions\cite{frixione}.
The bold curves refer to a charm mass value $m_c=1.5$ GeV and 
the renormalization scale $m_R=a\sqrt{m_c^2+p_{\perp}^2}$ with $a=1$.
From fig.\ref{fig:dsubs}a it is evident that the normalization 
of the theoretical prediction is significantly too low even for
extreme values $m_c=1.2$ GeV and $a=0.5$ (thin lines).
Furthermore the shape (fig.\ref{fig:dsubs}b) is 
inconsistent with the data which are much higher in the forward direction 
$\eta>0$. This behavior of the data could be due to an interaction between 
the color charges of the $c$-quark and the proton remnant (beam drag)\cite{beamdrag} 
which is not accounted for in the model. 
With the PYTHIA fragmentation the agreement is slightly better\cite{dsubstamp}.
In the same kinematic region a result of
$\sigma^{vis}_{D^*} = (9.17 \pm 0.35 (stat) ^{+0.40}_{-0.39} (syst))$ nb 
is obtained for $D^*$ production. The ratio
$\sigma^{vis}_{D_s}/\sigma^{vis}_{D^*}$ is determined as
$0.41\pm 0.07(syst)^{+0.03}_{-0.05}(stat)\pm 0.10(BR)$.
This result is compared to the ratio of fragmentation probabilities
$f(c\rightarrow D_s)/f(c \rightarrow D^*)=0.43\pm 0.04\pm 0.11$ as 
obtained from $e^+e^-$ experiments\cite{gladilin}.
The strangeness suppression factor
$\gamma_s$ is
the ratio of probabilities to create $s$ or ($u,d$) quarks during the 
fragmentation process. In simulations based on the Lund String Model 
\cite{pythia} $\gamma_s$ can be adjusted. For HERA an optimal 
value of $0.27\pm 0.05$ is found while for $e^+e^-$ 
experiments $0.26\pm 0.03$ had been obtained\cite{gladilin}.
While direct measurements
of the fragmentation function at HERA are not yet available, the
agreement shows a consistent picture of charm fragmentation which tends to
support the universality of charm fragmentation.

\subsection{$D^*$ cross sections at small $Q^2$}

Visible $D^*$ cross sections at small values of $Q^2$ 
have been measured by both 
H1\cite{osakaa} and ZEUS\cite{zeusf2cc} 
yielding
$\sigma_{ep\rightarrow eD^*X}^{\rm ZEUS}=
8.31 \pm 0.31 (stat) ^{+0.3}_{ -0.5}(syst)$ nb
and 
$\sigma_{ep\rightarrow eD^*X}^{\rm H1}
=8.37 \pm 0.41 (stat) ^{+1.11}_{-0.82} (syst) ^{+0.64}_{-0.39}
(theo) $ nb
for slightly different kinematic regions 
(ZEUS: $1<Q^2<600$ GeV$^2$, $y<0.7$, $1.5<p_{\perp}<15 $ GeV,$|\eta| <1.5$; 
 H1: $1<Q^2<100$ GeV$^2$, $0.05<y<0.7$, $p_{\perp}> 1.5 $ GeV,$|\eta| <1.5$).
The ZEUS data are shown in fig.\ref{fig:zeusdstarsmallq} 
in comparison with HVQDIS. The open
band corresponds to the standard Peterson fragmentation function.
To account for potential beam-drag effects the Peterson fragmentation 
is replaced by that of the RAPGAP simulation\cite{rapgap} (shaded band) leading to a somewhat improved
description of the $\eta^{D^*}$ distribution.
Fig.\ref{fig:h1dstarsmallq} shows the H1 data in comparison with
the result from HVQDIS (dark shaded) and from CASCADE (light shaded).
CASCADE describes the data fairly well in both normalization and shape
while HVQDIS is lower and fails to describe the shape of the 
$\eta_{D^*}$-distribution in particular in the forward direction.
At small values of $x$ CASCADE also provides 
a better description.

\section{Extraction of $F_2^{c\bar{c}}$}

\begin{figure}[t] 
\unitlength1.0cm
\begin{picture}(12.,8.)
\put(0.0,7.3){a)}
\put(8.0,7.3){b)}
\put(0.,0.){\epsfig{file=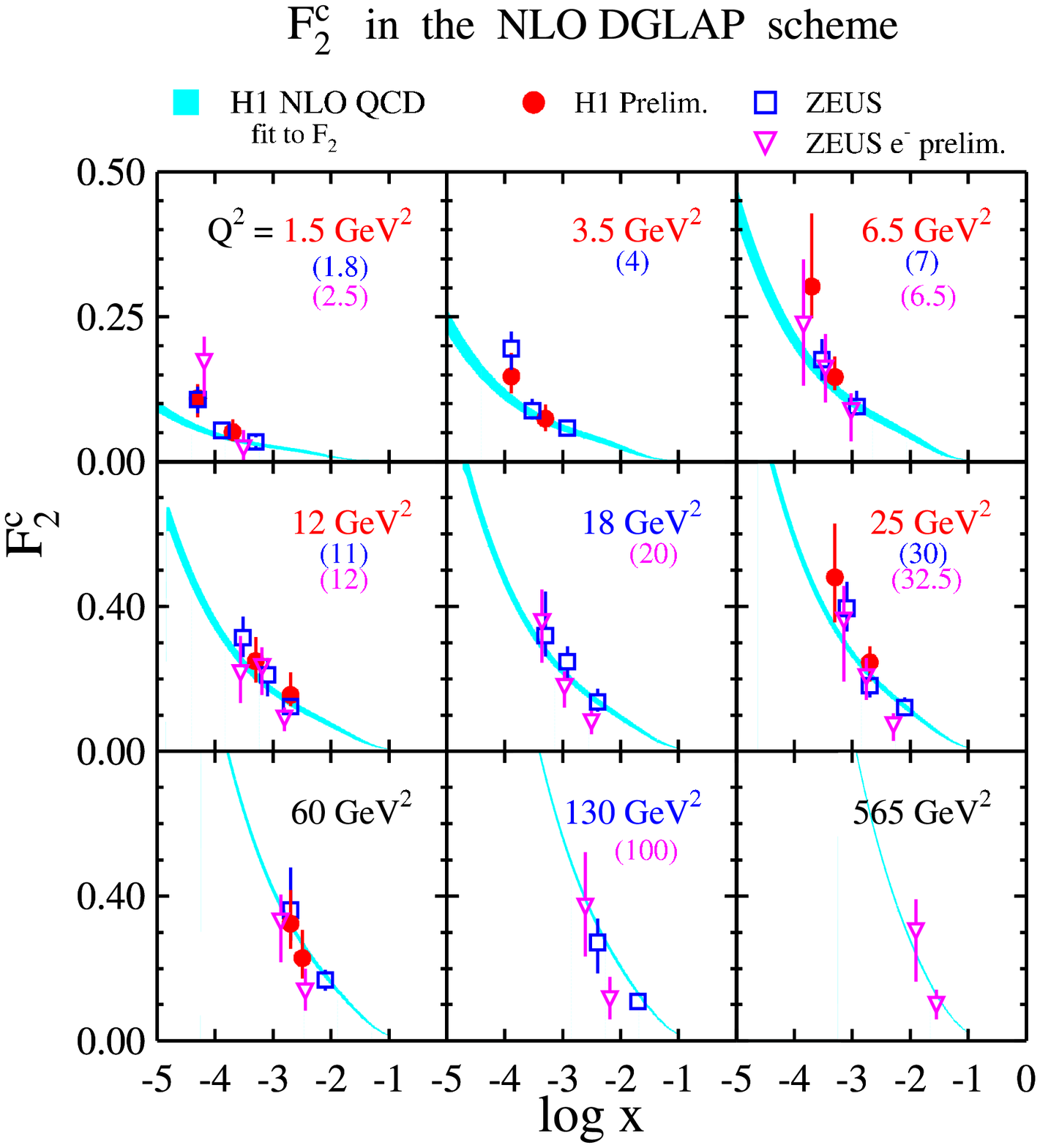,height=8.cm}}
\put(7.5,-0.1){\epsfig{file=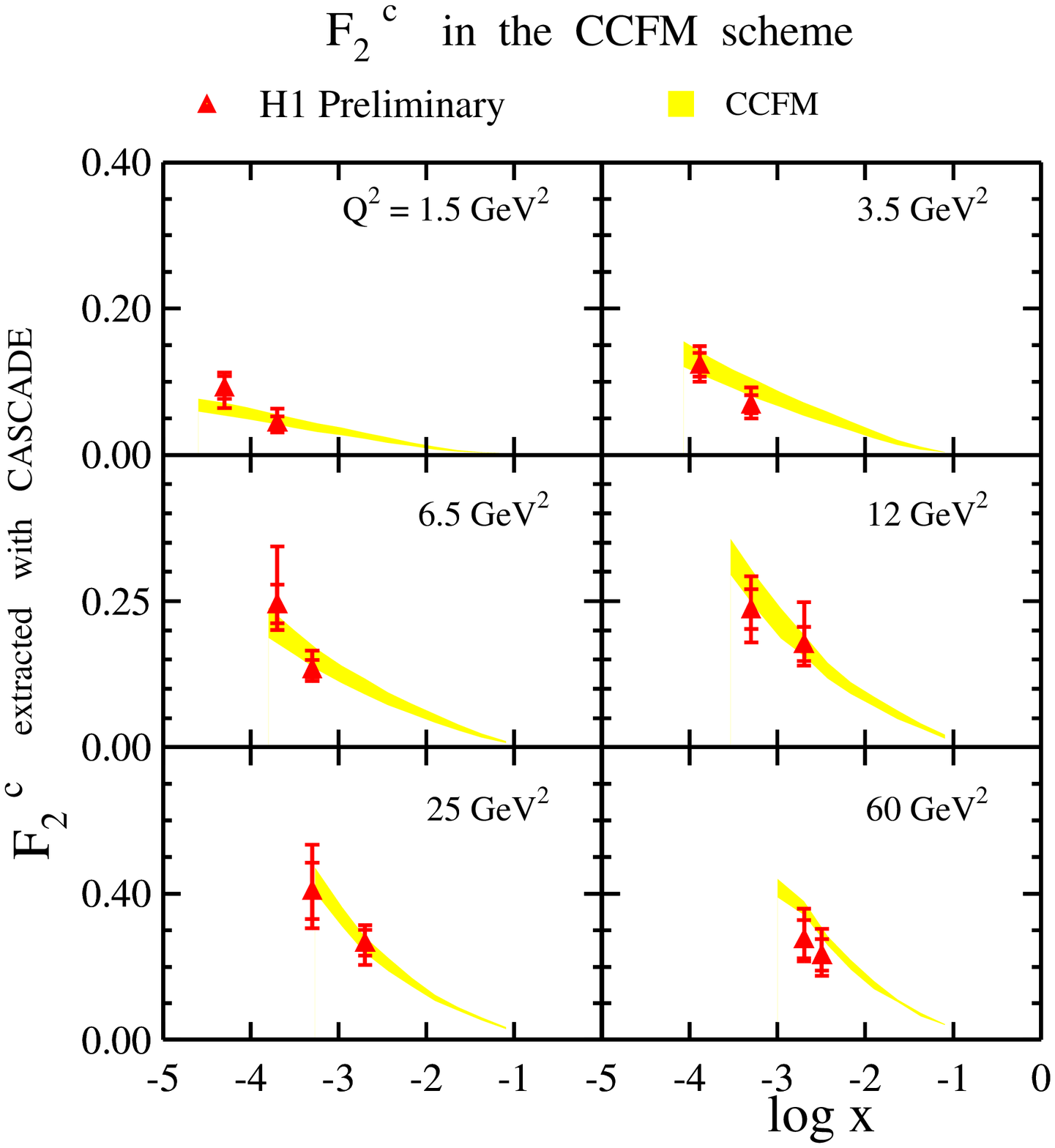,height=8.4cm}}
\end{picture}
\caption{\it The charm contribution to the proton structure function \f2cc
as derived a) using the HVQDIS program (H1, ZEUS) b) using the CCFM model as 
implemented in CASCADE (H1). Also shown are the predictions for 
\f2cc varying the charm quark mass between 1.3 and 1.5 GeV.
\label{fig:h1zeusf2cc} }
\end{figure}

The charm contribution to the proton structure function, $F_2^{c\bar{c}}$, 
is related to the charm cross section in $x$ and $Q^2$ as
$ \frac{d\sigma^{c\bar{c}}}{dxdQ^2} =
\frac{2\pi\alpha}{xQ^4}(1+(1-y)^2) \cdot F_2^{c\bar{c}}(x,Q^2).$
The contribution from $F_L$ is expected to be small and is neglected here.
\f2cc is extracted from the measured visible cross section in each bin of $x$ 
and $Q^2$ by extrapolation into the full phase space in $\eta$ and $p_{\perp}$
using the HVQDIS program. As shown in fig.\ref{fig:h1zeusf2cc} 
there is good agreement between H1 and ZEUS and, in general, 
with the prediction obtained from a NLO QCD fit to the inclusive 
H1 data. At the lowest $x$ and $Q^2$, a tendency towards a departure 
from the model is visible.
Alternatively the CCFM model as implemented in CASCADE 
can be used for the extrapolation. 
Fig.\ref{fig:h1zeusf2cc}b) shows the comparison between the H1 data
as extracted using CCFM and the CCFM expectation. 
There is good agreement also at the lowest values of $x$ and $Q^2$.
A comparison of \f2cc with the inclusive proton structure function $F_2$
shows that the contribution from charm to the inclusive cross section
is larger than 10\% everywhere in the region of $Q^2>1$ GeV$^2$ and $x<10^{-3}$ 
and can be as large as 25\% at large values of $Q^2$.

\section{Open Beauty Production}
\begin{figure}[t] 
\unitlength1.0cm
\begin{picture}(12.,5.5)
\put(0.5,4.8){a)}
\put(7.7,4.8){b)}
\put(0.5,0.){\epsfig{file=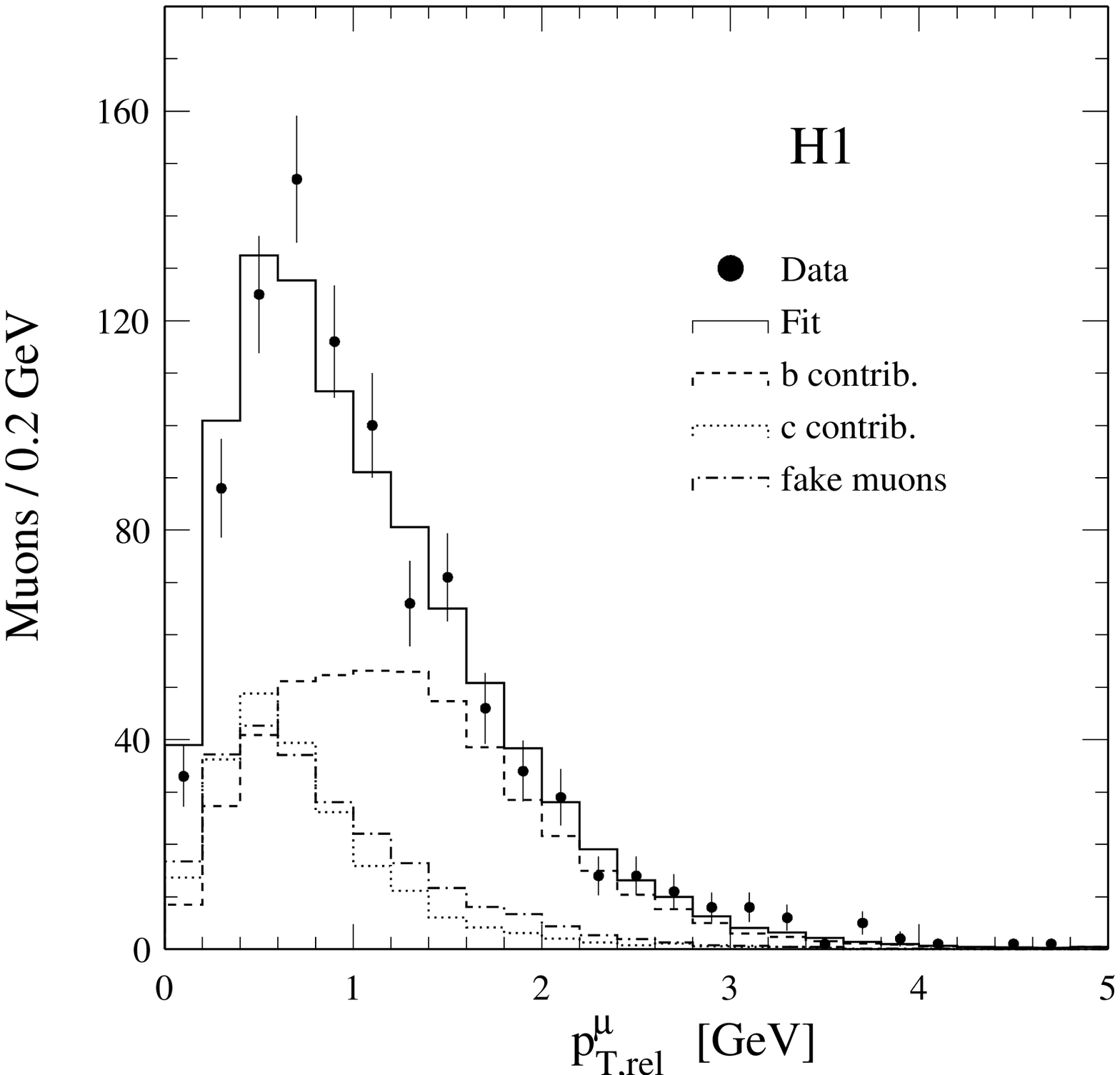,width=5.5cm}}
\put(7.5,-0.2){\epsfig{file=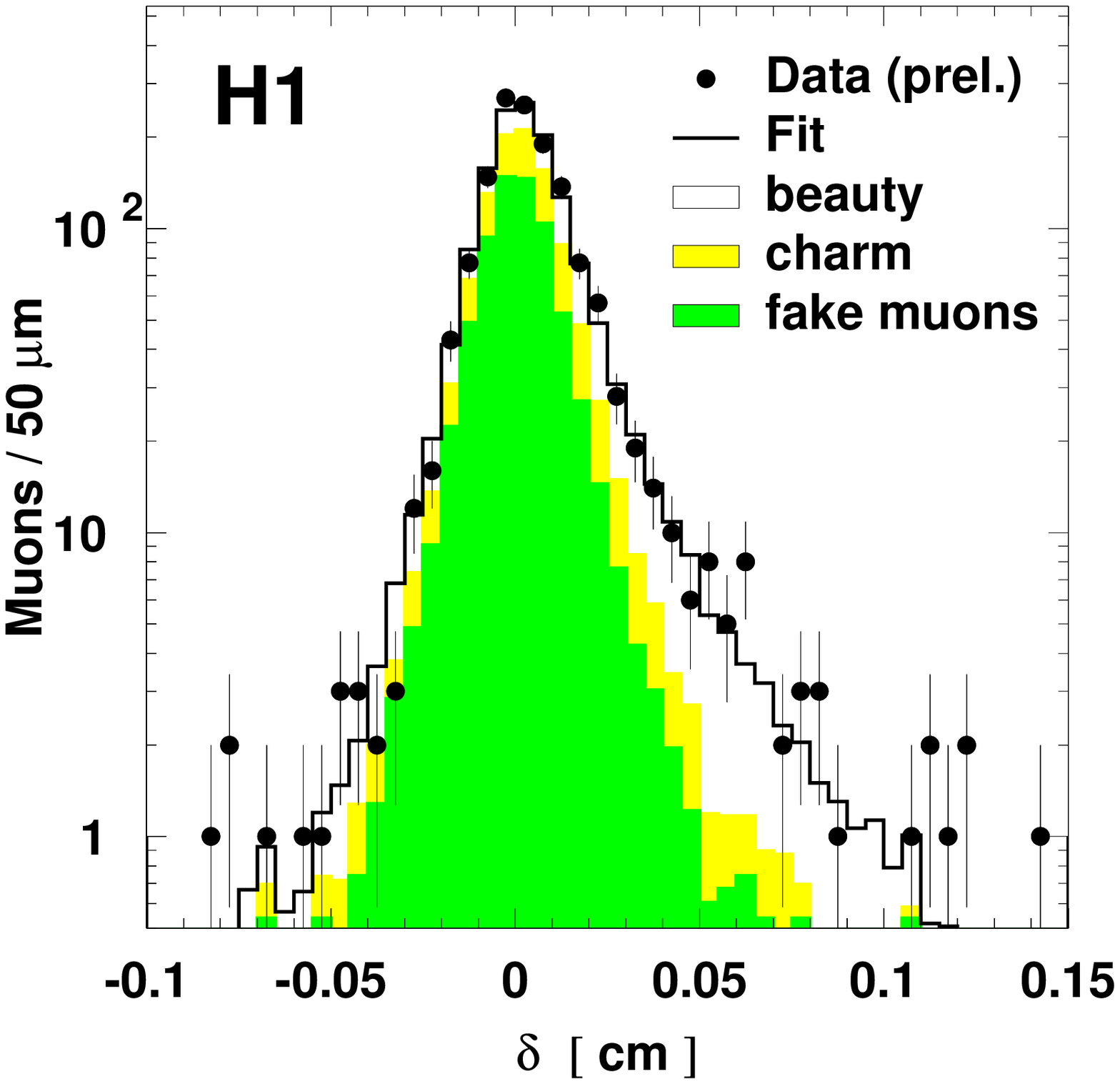,height=6.5cm,width=6.5cm}}
\end{picture}
\caption{\it 
a) $p^{\mu}_{\perp , rel}$ of the data and the fitted sum (solid line)
of the contributions from events with beauty, charm and fake muons.
b) Impact parameter and decomposition from the likelihood fit.
\label{fig:bptrel} }
\end{figure}

\begin{figure}[t] 
\unitlength1.0cm
\begin{picture}(12.,6.5)
\put(7.5,4.0){a)}
\put(0.,0.){\epsfig{file=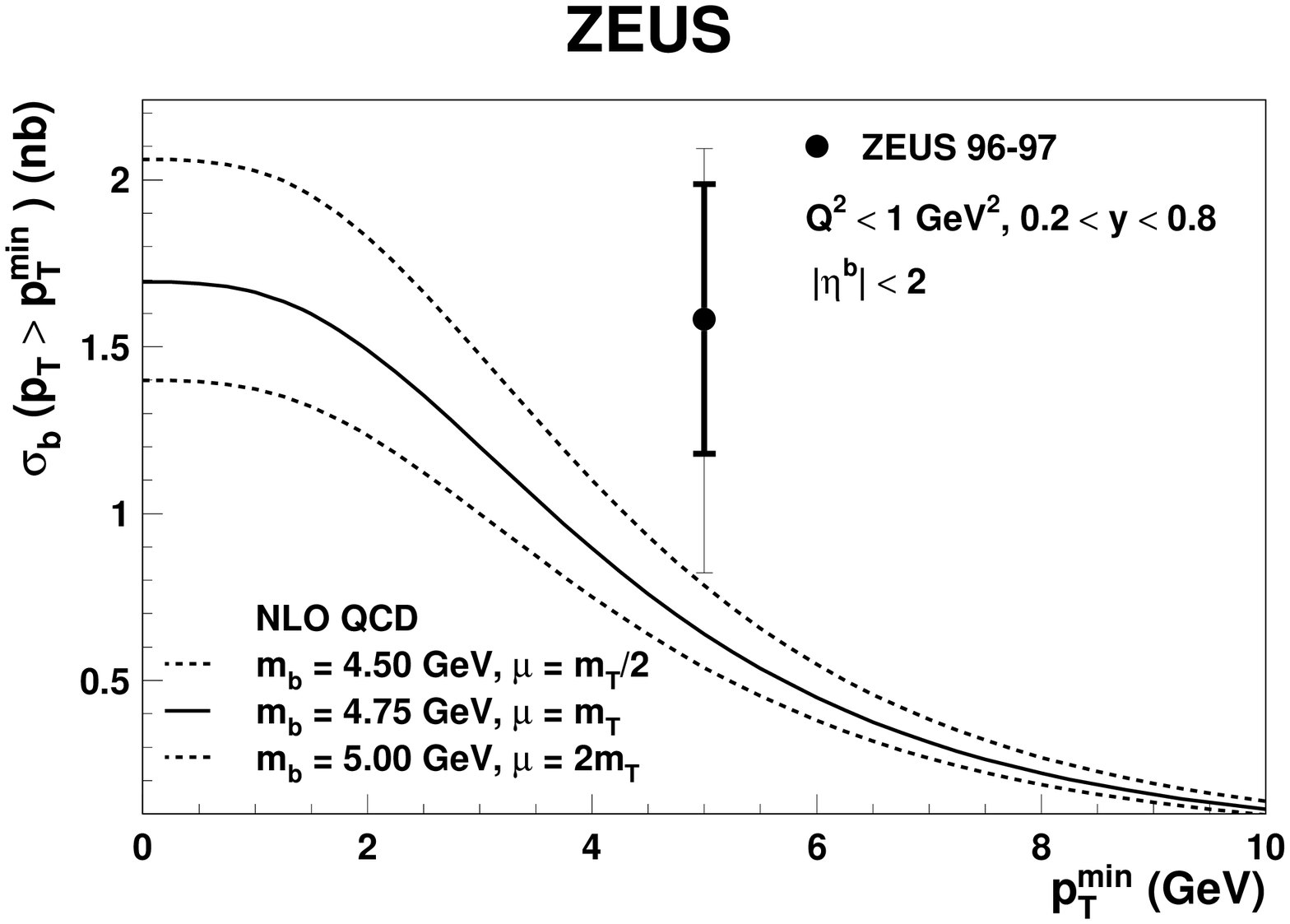,height=6.5cm}}
\put(8.7,0.){\epsfig{file=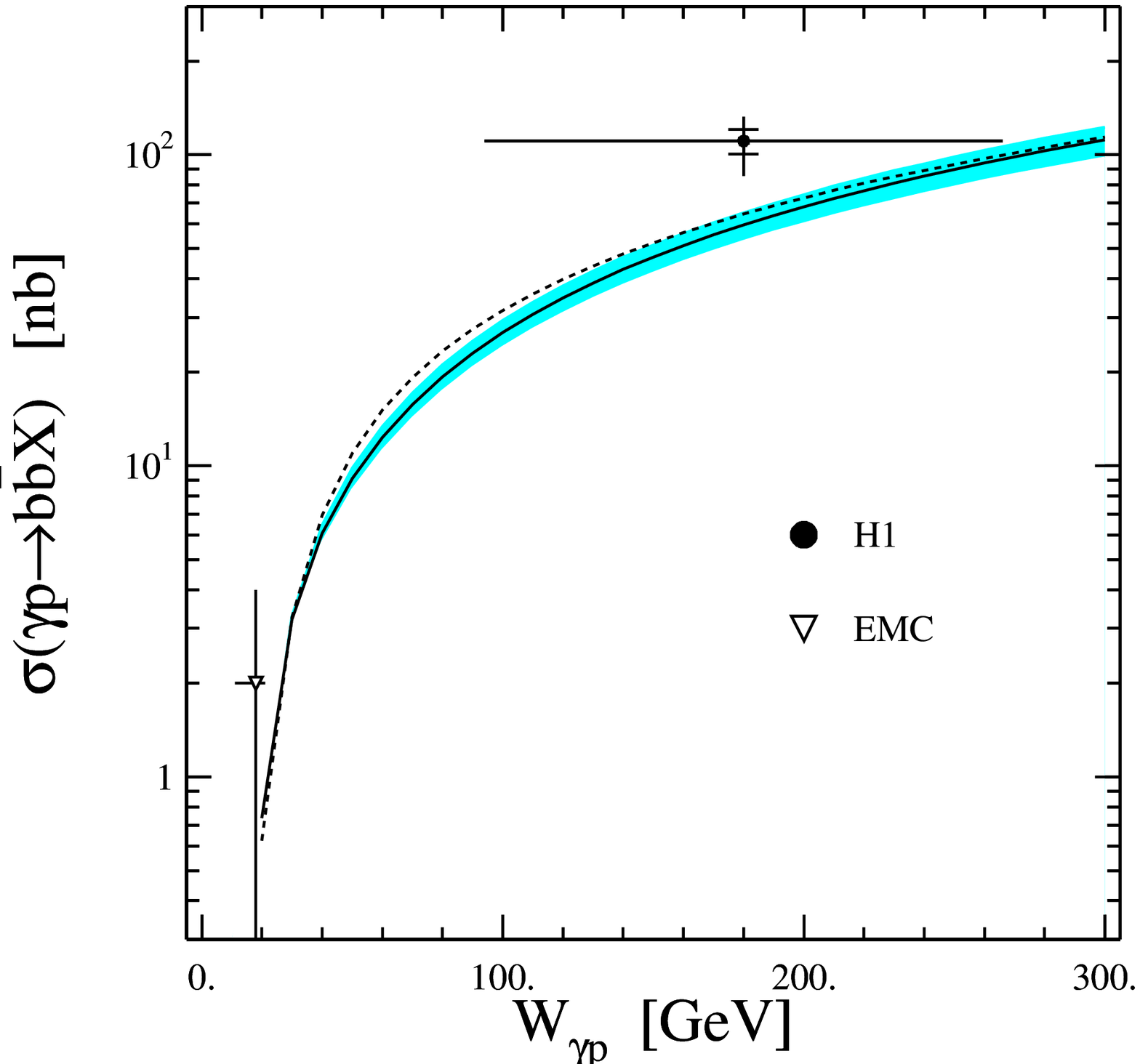,width=6.5cm}}
\put(14.0,4.0){b)}
\end{picture}
\caption{\it 
a) Extrapolated $b$ photoproduction cross section as measured by the ZEUS Experiment
for $p_T^b > p_T^{\rm min} = 5$ GeV, $0.2<y<0.8$ and $|\eta^b|<2 $ 
compared with NLO QCD predictions plotted as a function of $p_T^{\rm min}$. 
b) The total photoproduction cross section, $\sigma(\gamma p \rightarrow
b\bar{b} X)$ measured by H1. The horizontal error bar represents the range of the
measurement. The curve shows the expectation from the FMNR NLO calculation
using different structure functions.
 \label{fig:bcross} }
\end{figure}

In contrast to charm quarks, $b$-quarks are much heavier, 
thus providing a harder scale for perturbative calculations. At 
the same time the total $b\bar{b}$ cross section
is expected to be smaller than that of charm by a factor of 200.
The first measurements on the beauty production cross section 
in $ep$-scattering at HERA were performed using semi-leptonic 
decays\cite{bh1,bzeus,bzeustamp}.
In these analyses events with two or more jets are selected if they have
at least one identified muon (electron) with a $p_{\perp}$ of 2 GeV (1.6 GeV)
in the central region of the H1 (ZEUS) detector respectively. 
The lepton is assigned to one of the two jets 
of the event.
To distinguish $b$-events from those 
with charm or fake leptons and to extract the $b\bar{b}$ cross section
the transverse momentum $p_T^{rel}$ of the lepton relative to the jet is used
(fig.\ref{fig:bptrel}a). Both measurements (fig.\ref{fig:bcross}) are 
above the NLO expectations. 

\begin{figure}[t] 
\unitlength1.0cm
\begin{picture}(12.,6.5)
\put(0.5,5.5){a)}
\put(7.6,5.5){b)}
\put(0.5,0.){\epsfig{file=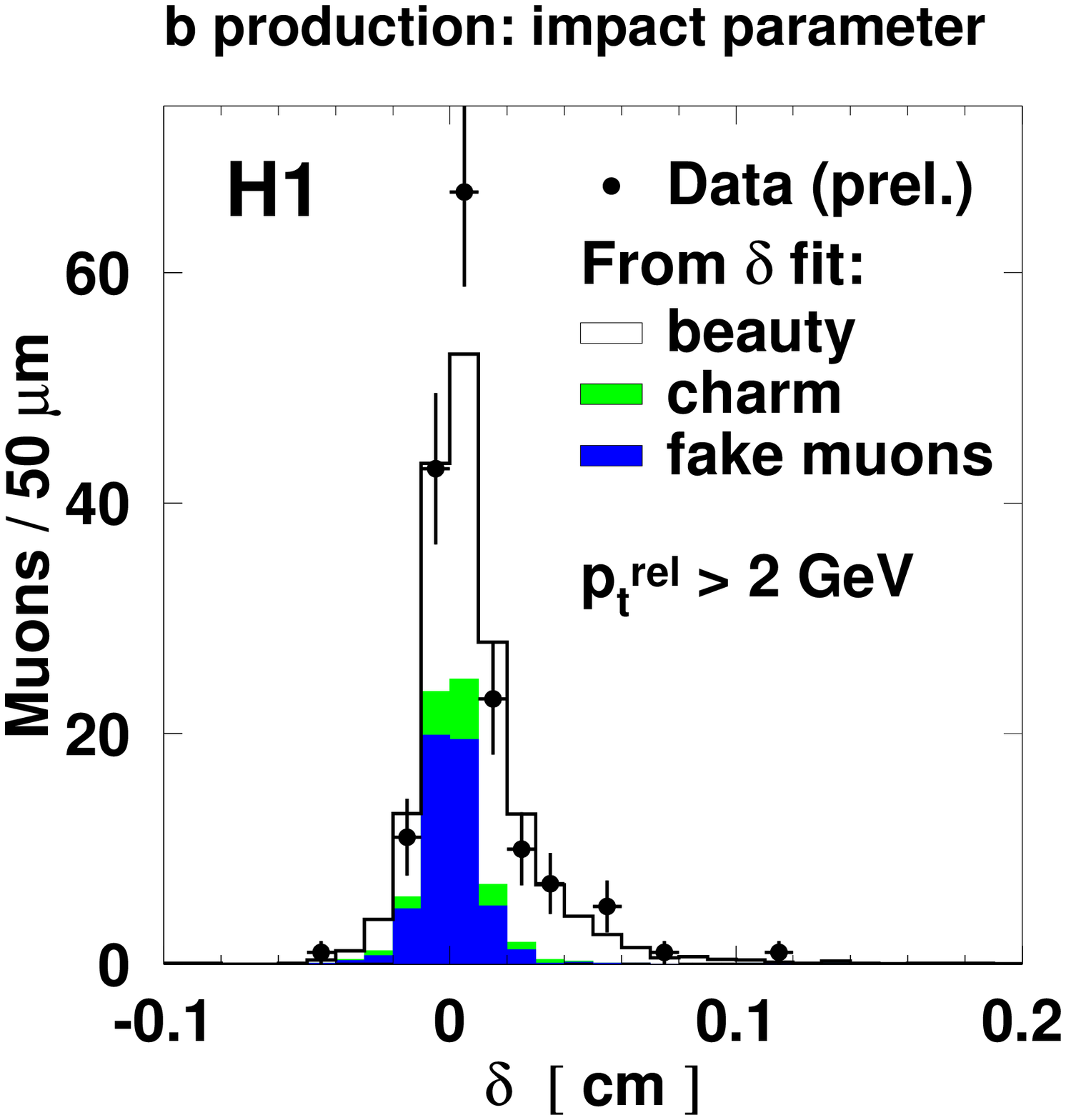,width=6.5cm}}
\put(7.5,0.){\epsfig{file=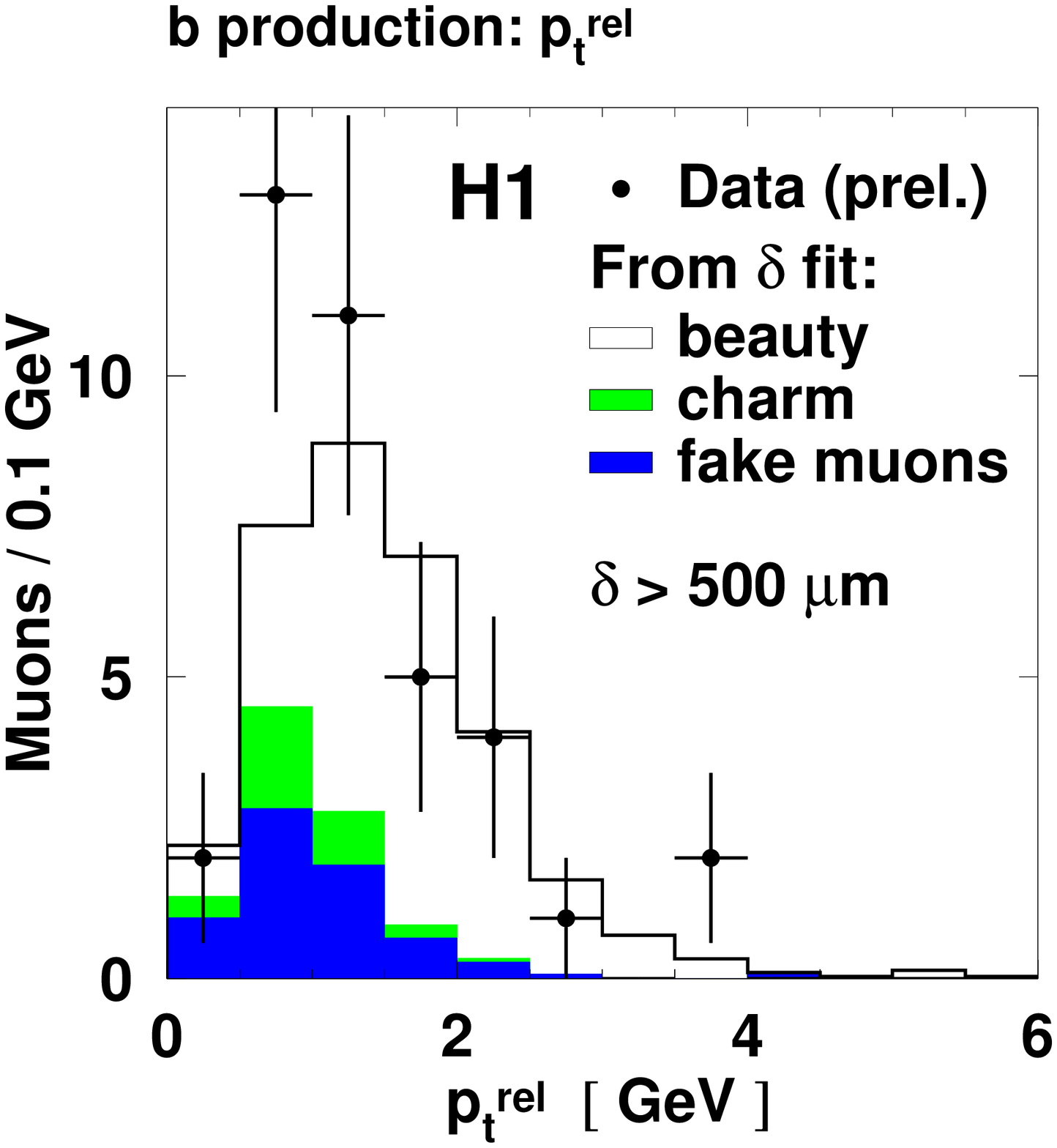,width=6.5cm}}
\end{picture}
\caption{\it 
a) Impact parameter distribution for muon candidates with 
$p_T^{rel}> 2$GeV and
b) $p_T^{rel}$ distribution for muon candidates with $\delta>500 \mu$m, 
with estimated contributions.
  \label{fig:bcomb} }
\end{figure}

Recently a new independent measurement has been presented by H1\cite{osakae}.
The analysis uses the central silicon vertex detector 
to improve the track resolution. Events with $b$-mesons are tagged
by their long lifetime using the impact parameter between the muon 
track and the beam in the transverse plane.
From the impact parameter distribution (fig.\ref{fig:bptrel}b)
the $b\bar{b}$-cross section is extracted. A log-likelihood fit
yields a value of $\sigma_{ep\rightarrow eb\bar{b}X}=(176 \pm 30(stat) \pm 29
(syst))$ pb which confirms the previous results.
To maximize the $b$-contribution in the event sample and to 
reduce the systematics the two methods are combined 
in a likelihood fit (fig.\ref{fig:bcomb}), yielding a cross section 
of $170\pm25$ pb. Like the previous measurements this result is 
somewhat higher than the NLO QCD expectation\cite{frixione} 
which gives $104 \pm 17$ pb.

\section{Summary}
Since 1992 HERA has been collecting a wealth of data
at continuously increasing luminosity.
New results with charm and beauty data have been presented here. 
In the charm sector precision tests of QCD are being 
performed by confronting the data in a variety of channels ($D^*$, $D_s$ and
semi-leptonic decays) with predictions from NLO calculations. 
The process of boson gluon fusion has been shown to be 
by far the dominant contribution to the production of charm quarks in 
the wide kinematic range between the photoproduction regime, 
$Q^2 \rightarrow 0$, and $Q^2 <$ 1000 GeV$^2$.
However, in photoproduction, presently, the models do not
fully reproduce the data in normalization and shape.
In the region of low $Q^2$ ($1<Q^2<600$GeV$^2$) the charm contribution
to the proton structure function \f2cc has been extracted.
The data have been compared to calculations using different parton
evolution schemes and there are indications that the CCFM equation
gives a better description of the data in the region of the 
smallest values of $x$.
In the beauty sector a new 
systematically independent measurement using
the lifetime tag (impact parameter) of semi-muonic events
has been presented.
The result confirms the previous measurements from H1 and ZEUS
which are somewhat larger than expected by NLO calculations.

\section{Acknowledgments}
I wish to thank the organizers for a very pleasant 
and fruitful meeting and my 
colleagues at ZEUS and H1 for supplying their data and the discussions.


\begin{thebibliography}{99}
\bibitem{hvqdis}B.W.\,Harris, J.\,Smith, Phys.\,Rev.\,D57 (1998) 2806.
\bibitem{dglap} V.\,N.\,Gribov and L.\,N.\,Lipatov,
Sov.\,J.\,Nucl.\,Phys.\, 15 (1972) 438;
L.\,N.\,Lipatov, Sov.\,J.\,Nucl.\,Phys.\, 20 (1975) 96.;
G.\,Altarelli and G.\,Parisi, Nucl.\,Phys. B126 (1977) 298.
\bibitem{ccfm}M.\,Ciafaloni, Nucl.\,Phys.\, B296 (1988) 49; S.\,Catani, F.\,Fiorani
and G.\,Marchesini, Phys.\,Lett.\, B234 (1990) 339; Nucl.\,Phys.\, B336 (1990) 18.
\bibitem{cascade}H.\,Jung, Nucl.\,Phys.\, B (Proc.\,Suppl.)
79 (1999) 429; hep-ph/9905554.
\bibitem{osakaa} 
H1 Coll.,~contr.~paper\footnote{Contributed~papers~to~International~Conference~on~High~Energy
Physics,~ICHEP2000,~Osaka,~July 2000.
{\tt http://www-h1.desy.de/h1/www/publications/conf/list.ICHEP2000.html \rm and}
{\tt http://www-zeus.desy.de/physics/phch/conf/osaka\_paper.html}
} to ICHEP2000, Osaka, 313.
\bibitem{osakab} ZEUS Coll., contr.\,paper to ICHEP2000, Osaka, 449.
\bibitem{zeusf2cc} ZEUS Coll., J.\,Breitweg et al., Eur.\,Phys.\,J. C12 (2000) 35.
\bibitem{dsubs}ZEUS Coll., J.\,Breitweg et al., Phys.\,Lett.\,B481 (2000) 213.
\bibitem{osakac} ZEUS Coll., contr.\,paper to ICHEP2000, Osaka, 447.
\bibitem{bh1}H1 Coll., C.\,Adloff et al., Phys.\,Lett.\,B467 (1999) 156.
\bibitem{bzeustamp} ZEUS Coll., contr.\,paper to IECHEP1999, Tampere, 498.
\bibitem{bzeus}ZEUS Coll., J.\,Breitweg et al., DESY 00-166, 
accepted by Eur.\,Phys.\,J.\, C.
\bibitem{osakae} H1 Coll. contr.\,paper to ICHEP2000, Osaka, 311.
\bibitem{osakad} ZEUS Coll., contr.\,paper to ICHEP2000, Osaka, 448.
\bibitem{zeusqcd}ZEUS Coll., J.\,Breitweg et al., Eur.\,Phys.\,J.\, C7
(1999) 609.
\bibitem{grv98} M.\,Gl\"uck, E.\,Reya, A.\,Vogt, Eur.\,Phys.\,J.\, C5 (1998) 461.
\bibitem{cteq5} H.L.\,Lai et al., Eur.\,Phys.\,J.\, C12 (2000) 375.
\bibitem{peterson} C.\,Peterson et al., Phys.\,Rev.\, D27 (1983) 105.
\bibitem{dstargp} ZEUS Collab., J.\,Breitweg et al., 
Eur.\,Phys.\,J.\, C6 (1999) 67;\\
H1 Collab., C.\,Adloff et al., Z.\,Phys. C72 (1996) 593.
\bibitem{frixione}S.\,Frixione et al., Phys.\,Lett.  B348, 633 (1995);
 Nucl. Phys.  B454, 3 (1995).
\bibitem{beamdrag} E.\,Norrbin, T.Sj\"ostrand, hep-ph/9905493.
Proc. of DESY Workshop "Monte Carlo Generators for HERA Physics".
\bibitem{dsubstamp} ZEUS Coll., contr.\,paper to IECHEP1999, Tampere, 525.
\bibitem{gladilin} L. Gladilin, hep-ex/9912064, 1999.
\bibitem{pythia} T.Sj\"ostrand, Comp.\,Phys.\,Commun.\, 82 (1994) 74.
\bibitem{rapgap} H.\,Jung,  Comp. Phys. Commun.  86, 147 (1995).

\end{thebibliography}
\end{document}